\documentclass[11pt, a4paper]{article}
\usepackage{jheparxiv}
\usepackage[latin1]{inputenc}
\usepackage{amsmath}
\usepackage{amsfonts}
\usepackage{amssymb}
\usepackage{latexsym}
\usepackage{mathrsfs}
\usepackage{graphicx}
\usepackage{color}
\usepackage{slashed}
\usepackage{twistor}

\renewcommand{\d}{\mathrm{d}}

\subheader{\hfill \texttt{DAMTP-2015-19}}

\title{Perturbative gauge theory at null infinity}

\author{Tim Adamo and Eduardo Casali}

\affiliation{Department of Applied Mathematics \& Theoretical Physics \\
        University of Cambridge \\
        Wilberforce Road \\
        Cambridge CB3 0WA, United Kingdom}

\emailAdd{[t.adamo, e.casali]@damtp.cam.ac.uk}

\abstract{We describe a theory living on the null conformal boundary $\scri$ of four-dimensional Minkowski space, whose states include the radiative modes of Yang-Mills theory. The action of a Kac-Moody symmetry algebra on the correlators of these states leads to a Ward identity for asymptotic `large' gauge transformations which is equivalent to the soft gluon theorem.  The subleading soft gluon behavior is also obtained from a Ward identity for charges acting as vector fields on the sphere of null generators of $\scri$.  Correlation functions of the Yang-Mills states are shown to produce the full classical S-matrix of Yang-Mills theory. The model contains additional states arising from non-unitary gravitational degrees of freedom, indicating a relationship with the twistor-string of Berkovits \& Witten.}

\notoc 

\begin{document}
 
\maketitle

\vfill

\pagebreak

\section{Introduction}

In asymptotically flat space-times, the S-matrix dictates how asymptotic, non-interacting states in the far past evolve through the bulk to asymptotic states in the far future. For massless particles, the natural habitat for these asymptotic states is the conformal boundary $\scri=\scri^{-}\cup\scri^{+}$, where $\scri^{-}$ and $\scri^{+}$ are the past and future null boundaries, respectively~\cite{Penrose:1962ij,Penrose:1965am}. In this sense, the massless S-matrix is holographic almost by definition. However, standard methods for its computation are decidedly not: techniques based on Feynman diagrams explicitly propagate asymptotic states through the bulk.

As in the AdS/CFT correspondence, we expect asymptotic symmetries to play an important role in determining any holographic description of a massless field theory coupled to gravity in the bulk. In four space-time dimensions, the asymptotic symmetry group of gravity is the BMS group~\cite{Bondi:1962px,Sachs:1962wk}. Recently, Strominger and collaborators have shown that, in a certain class of space-times which are natural from a scattering point-of-view, the BMS group is a symmetry of the gravitational S-matrix~\cite{Strominger:2013jfa,He:2014laa}. The associated Ward identities include Weinberg's soft graviton theorem~\cite{Weinberg:1965nx}, proving that the asymptotic symmetry group encodes universal information about quantum gravity in the bulk. Furthermore, extending the BMS group to include certain \emph{local} asymptotic symmetries~\cite{Barnich:2009se,Barnich:2011mi} leads to Ward identities encoding subleading soft graviton behavior~\cite{White:2011yy,Cachazo:2014fwa,Kapec:2014opa}, with limited quantum corrections~\cite{Bern:2014oka,Cachazo:2014dia,He:2014bga,Bianchi:2014gla,Broedel:2014bza}.

Naturally, Yang-Mills theory coupled to gravity in the bulk should also have a description in terms of asymptotic degrees of freedom. The nature of the asymptotic \emph{gauge} symmetries depends on the boundary conditions imposed; a natural choice corresponds to requiring all asymptotic energy and charge fluxes of the gauge field to be finite. Under these conditions, gauge theory has `large' asymptotic gauge symmetries depending only on the conformal sphere at $\scri$, in addition to global gauge transformations~\cite{Strominger:2013lka,Barnich:2013sxa}. The Ward identity for these `large' gauge transformations is associated with a Kac-Moody symmetry on the sphere of null generators of $\scri$, and has been shown to encode the soft gluon theorem for both abelian and non-abelian gauge groups~\cite{He:2014cra,Mohd:2014oja,He:2015zea}.

There is a subleading soft gluon theorem analogous to the subleading soft graviton theorem~\cite{Casali:2014xpa}. In the abelian setting of QED, this was first found in the 1950s by Low~\cite{Low:1954kd,Low:1958sn}, and has now also been derived from an asymptotic symmetry perspective~\cite{Lysov:2014csa}. Surprisingly, the associated symmetry is \emph{not} a gauge symmetry, but acts  geometrically as a vector field on the sphere of null generators at $\scri$. It seems natural to expect that this is also true for non-abelian gauge theory. As in gravity, this subleading soft theorem is tightly constrained -- and with minimal additional assumptions, determined -- by Poincar\'e and gauge invariance, giving further credence to its interpretation as a universal feature of gauge theory~\cite{Larkoski:2014hta,Broedel:2014fsa,Bern:2014vva,White:2014qia}.

\medskip

In~\cite{Adamo:2014yya}, a dynamical realization for these ideas was given at the semi-classical level for gravity and the BMS group. In this paper, we provide a dynamic realization at the semi-classical level for \emph{gauge theory}. We define a worldsheet CFT with target space the (complexified) conformal boundary of four-dimensional Minkowski space, whose states include all radiative degrees of freedom for $\cN=4$ super-Yang-Mills. A current algebra induces a natural Kac-Moody action on correlators; associated charges include global as well as `large' asymptotic gauge transformations. The action of these charges leads to Ward identities encoding the soft gluon theorem. We also find charges implementing rotations on the sphere of null generators at $\scri$, whose Ward identity gives the subleading soft gluon theorem. Furthermore, correlation functions of the model give the entire tree-level S-matrix of gauge theory.

Of course, one should expect an asymptotic description of gauge theory to be coupled to gravity in some way, and indeed our model contains gravitational states. However, we argue that these gravitational states are non-unitary, corresponding to a theory with fourth-order equations of motion. These states are decoupled from the single-trace color structure of our semi-classical analysis, and so do not affect any Ward identities at tree-level, but they do mediate multi-trace interactions.

As in~\cite{Adamo:2014yya}, we emphasize that this model should \emph{not} be viewed as a truly holographic description of Yang-Mills coupled to a certain gravitational theory in the bulk. It should instead be viewed as a perturbative description of such a model which holds in a regime where classical gauge theory is a valid bulk description. An alternative CFT realization of leading and subleading soft theorems is provided by the ambitwistor approach~\cite{Mason:2013sva,Geyer:2014fka}. In these models the presentation of asymptotic data is not in terms of radiative modes, but rather field perturbations; soft theorems are generated by insertions of soft vertex operators~\cite{Geyer:2014lca,Lipstein:2015rxa}. 

\medskip

Section \ref{sec:asymp} begins with a review of the asymptotic geometry, using the machinery employed in~\cite{Adamo:2014yya}. We also discuss how the radiative degrees of freedom for the bulk gauge field are encoded asymptotically. The model is defined in Section \ref{sec:model}, along with its BRST charge and spectrum of states. In Section \ref{sec:KMsym}, we realize the asymptotic symmetries of the gauge field as Kac-Moody currents in the model and show how they act on correlators, leading to the soft gluon theorem. Currents and Ward identities for the subleading soft gluon theorem are also given. Section \ref{sec:amps} shows that correlators of gauge-theoretic states give the complete tree-level S-matrix of $\cN=4$ super-Yang-Mills, and also discusses the gravitational degrees of freedom in the model. Section \ref{sec:con} concludes with some comments on the relationship between our model and the original twistor-string, as well as potential generalizations.


\section{Asymptotic Geometry and Gauge Fields}
\label{sec:asymp}

In four-dimensions, the conformal boundary of (conformally compactified) Minkowski space is a null hypersurface $\scri$, composed of disjoint past and future boundaries  $\scri=\scri^{-}\cup\scri^{+}$, each having the topology of a light cone $\scri^{\pm}\cong\R\times S^2$~\cite{Penrose:1962ij,Penrose:1965am}.  For a massless field theory living in Minkowski space, the S-matrix describes scattering which propagates asymptotic states from $\scri^{-}$ to $\scri^{+}$ (in Lorentzian signature). The analyticity of Minkowski space allows us to consider a single \emph{complex} manifold $\scri_{\C}$, which we take to be the complexification of either $\scri^{\pm}$, rather than two disjoint real boundaries boundaries (\textit{c.f.}, \cite{Penrose:1986}). Crossing symmetry of the S-matrix implies that massless scattering amplitudes are analytically continued to functions of asymptotic states on $\scri_{\C}$.

Usually, the complex three-manifold $\scri_{\C}$ is charted with coordinates $(u,\zeta,\tilde{\zeta})$, where $u$ is a coordinate along the null generators and the complex stereographic coordinates $(\zeta,\tilde{\zeta})$ parametrize the complexified two-sphere of null directions (\textit{c.f.}, \cite{Newman:1981fn,Adamo:2009vu}). However, we follow~\cite{Adamo:2014yya} and adopt a \emph{projective} presentation of $\scri_{\C}$. To do this, we view the complexified space of null generators as the product of two Riemann spheres $\CP^{1}\times\CP^{1}$, each charted with homogeneous coordinates $\lambda_{\alpha}=(\lambda_{0},\lambda_{1})$ and $\tilde{\lambda}_{\dot\alpha}=(\tilde{\lambda}_{\dot0},\tilde{\lambda}_{\dot1})$, respectively. This gives projective coordinates $(u,\lambda,\tilde{\lambda})$ on $\scri_{\C}$, with the scaling equivalence~\cite{Eastwood:1982}
\begin{equation*}
 (u,\lambda,\tilde{\lambda})\sim (r\tilde{r}u, r\lambda, \tilde{r}\tilde{\lambda}), \qquad r,\tilde{r}\in\C^{*}\,.
\end{equation*}
This description is equivalent to viewing $\scri_{\C}$ as the total space of the line bundle
\be\label{sc1}
\cO(1,1)\rightarrow\CP^{1}\times\CP^{1}\,,
\ee
where $\cO(m,n)$ is the line bundle of complex functions on $\CP^{1}\times\CP^{1}$ which are homogeneous of weight $m$ in $\lambda$ and $n$ in $\tilde{\lambda}$. As usual, the real $\scri^{\pm}$ is recovered by imposing reality conditions $u=\bar{u}$, $\tilde{\lambda}_{\dot\alpha}=\overline{\lambda_{\alpha}}$.

One advantage of this formalism is that it naturally allows for the incorporation of supersymmetry in a parity symmetric way. For $\cN=2p$ supersymmetry, one simply extends the complexified space of null generators to $\CP^{1|p}\times\CP^{1|p}$, and the homogeneous coordinates to $\lambda_{A}=(\lambda_{\alpha},\eta_{a})$, $\tilde{\lambda}_{\dot{A}}=(\tilde{\lambda}_{\dot\alpha},\tilde{\eta}_{\dot{a}})$, where $\eta_{a}$, $\tilde{\eta}_{\dot{a}}$ are anti-commuting and $a,\dot{a}=1\ldots,p$. For this paper, we set $p=2$, corresponding to a parity symmetric presentation of $\cN=4$ supersymmetry. As we will see, this is adapted to the description of maximally supersymmetric Yang-Mills theory propagating in the bulk Minkowski space. 

The total space of $\cO(1,1)\rightarrow\CP^{1|2}\times\CP^{1|2}$ will also be denoted $\scri_{\C}$, with the distinction being clear from the context. Any functions on $\scri_{\C}$ can then be expanded polynomially in $\eta,\tilde{\eta}$ with the coefficients functions on the underlying bosonic conformal boundary. 

\medskip

Suppose that Yang-Mills theory (with gauge group $G$) describes gauge-theoretic massless fields propagating in the bulk Minkowski space. Following~\cite{Strominger:2013lka}, we impose boundary conditions on the gauge field such that the charge and energy flux through any subset of $\scri$ is finite. With these boundary conditions, it is easy to show that the asymptotic radiative degrees of freedom of the gauge field are controlled by a single function on $\scri$, taking values in the Lie algebra $\mathfrak{g}$ of the gauge group.

Analytically continuing the gauge field to complexified Minkowski space, the function controlling the asymptotic radiative information becomes a $\mathfrak{g}$-valued function on $\scri_{\C}$, taking values in $\cO(-1,1)$. We denote this function by $\cA^{0}(u,\lambda,\tilde{\lambda})$, suppressing gauge indices. The energy flux of the gauge field from the interior of Minkowski space through $\scri_{\C}$ is encoded by the \emph{broadcasting function} (\textit{c.f.}, \cite{Vanderburg:1969,Exton:1969im,Ashtekar:1981bq})
\be\label{broad}
\cF(u,\lambda,\tilde{\lambda})=\frac{\partial \cA^{0}}{\partial u}\,,
\ee
taking values in $\cO(-2,0)\otimes\mathfrak{g}$.\footnote{In the case $G=\U(1)$, the broadcasting function is the Newman-Penrose coefficient for the asymptotic Maxwell field $\phi^{0}_{2}$~\cite{Newman:1961qr}.} This broadcasting function is the gauge theoretic version of the Bondi news function~\cite{Bondi:1962px}, which encodes the radiative information of the gravitational field in asymptotically flat space-times.

Since the S-matrix specifies scattering data in terms of asymptotic insertions on $\scri_{\C}$, we expect this data to be given in by insertions of the broadcasting function. In a momentum eigenstate representation, the on-shell (complexified) four-momenta of these insertions are given by specifying a point on the space of null generators $(\lambda,\tilde{\lambda})$ and a scale or `frequency' $\omega$:
\be\label{4mom}
p_{\alpha\dot\alpha}=\omega\,\lambda_{\alpha}\,\tilde{\lambda}_{\dot\alpha}\,,
\ee
up to $(\omega,\lambda,\tilde{\lambda})\sim(r^{-1}\tilde{r}^{-1}\omega, r\lambda, \tilde{r}\tilde{\lambda})$. The frequency $\omega$ is naturally conjugate to the coordinate $u$, so broadcasting insertions of this type should be presented as plane waves $\e^{i\omega u}$ localized along a particular generator $(\lambda,\tilde{\lambda})$ of $\scri_{\C}$.


\section{The Model}
\label{sec:model}

As in the gravitational case studied in~\cite{Adamo:2014yya}, our model will be a CFT on the Riemann sphere $\Sigma$ governing holomorphic maps from $\Sigma$ to $\scri_{\C}$. We take $\scri_{\C}$ to be compatible with $\cN=4$ supersymmetry, being the total space of $\cO(1,1)\rightarrow\CP^{1|2}\times\CP^{1|2}$ as described above. The $\GL(1,\C)\times\GL(1,\C)$ scaling built into the projective description of $\scri_{\C}$ is given on the Riemann sphere in terms of two line bundles $\cL,\tilde{\cL}\rightarrow \Sigma$, of degree $d,\tilde{d}\geq0$. Following~\cite{Adamo:2014yya}, the coordinates on $\scri_{\C}$ are given by worldsheet fields
\begin{equation*}
 u\in\Omega^{0}(\Sigma,\cL\otimes\tilde{\cL})\,, \qquad \lambda_{A}\in\Omega^{0}(\Sigma, \C^{2|2}\otimes\cL)\,, \qquad \tilde{\lambda}_{\dot{A}}\in\Omega^{0}(\Sigma,\C^{2|2}\otimes\tilde{\cL})\,,
\end{equation*}
with the chiral, first-order action:
\be\label{S1}
S_{1}=\frac{1}{2\pi}\int_{\Sigma} w\,\dbar u +\nu^{A}\,\dbar\lambda_{A}+\tilde{\nu}^{\dot{A}}\,\dbar\tilde{\lambda}_{\dot{A}}\,.
\ee
The conformal weight $(1,0)$ fields $\{w,\nu^{A},\tilde{\nu}^{\dot{A}}\}$ are conjugate to the coordinates on $\scri_{\C}$. 

To this, we add the action for a worldsheet current algebra corresponding to the space-time gauge group $G$, $S_{C}$. As usual, this can be realized explicitly in terms of free fermions on the worldsheet. Additionally, we gauge the two $\GL(1,\C)$ symmetries associated with the line bundles $\cL$ and $\tilde{\cL}$ as well as two-dimensional gravity on the worldsheet.  This results in a ghost action
\be\label{S2}
S_{2}=\frac{1}{2\pi}\int_{\Sigma}\mathrm{m}\,\dbar\mathrm{n}+\tilde{\mathrm{m}}\,\dbar\tilde{\mathrm{n}}+b\,\dbar c\,,
\ee
where all ghost fields have fermionic statistics and $\mathrm{n},\tilde{\mathrm{n}}\in\Omega^{0}(\Sigma)$, $c\in\Omega^{0}(\Sigma, T_{\Sigma})$. Finally, we add an additional set of fermionic fields
\begin{equation*}
 \chi\in\Omega^{0}(\Sigma, \cL\otimes\tilde{\cL})\,, \qquad \xi\in\Omega^{0}(\Sigma, K_{\Sigma}\otimes(\cL\otimes\tilde{\cL})^{-1})\,,
\end{equation*}
with action
\be\label{S3}
S_{3}=\frac{1}{2\pi}\int_{\Sigma}\xi\,\dbar\chi\,.
\ee
The role of these fields (anomaly cancellation and defining correlation functions) will be revealed shortly.

This leaves us with the full action $S=S_{1}+S_{2}+S_{3}+S_{C}$ and a BRST charge implementing scale and conformal invariance on $\Sigma$:
\be\label{BRST}
Q=\oint c\,T-\mathrm{n}\left(w\,u+\nu^{A}\,\lambda_{A}+\chi\,\xi\right)-\tilde{\mathrm{n}}\left(w\,u+\tilde{\nu}^{\dot{A}}\,\tilde{\lambda}_{\dot{A}}+\chi\,\xi\right)\,,
\ee
where $T$ is the (normal-ordered) holomorphic stress tensor of the action. It is easy to see that the anomalies associated with $\cL,\tilde{\cL}$ vanish, and the theory has vanishing conformal anomaly provided the worldsheet current algebra contributes $+30$ to the central charge. Since we will only be concerned with the Riemann sphere ($\Sigma\cong\CP^1$), the central charge constraint can be relaxed.

\medskip

Vertex operators of the model sit inside the cohomology of the BRST charge $Q$. The structure of \eqref{BRST} makes clear that these vertex operators should be functions un-charged under $\cL\times\tilde{\cL}$ of conformal weight zero. Such vertex operators fall into two broad classes: gauge theoretic and gravitational. We are primarily interested in the former, which are given by
\be\label{gtvo}
c\, \mathrm{tr}\left(O(u,\lambda,\tilde{\lambda})\; j\right)\,,
\ee
where $O(u,\lambda,\tilde{\lambda})$ is a homogeneous function on $\scri_{\C}$ taking values in $\mathfrak{g}$ (the Lie algebra of the gauge group) and $j$ is the conformal weight $(1,0)$ Kac-Moody current provided by the worldsheet current algebra. The trace is paired with a $c$-ghost, which absorbs the conformal weight of the current $j$ and inserts the operator at a fixed point on $\Sigma$ with respect to the worldsheet conformal structure. As usual, the current obeys
\be\label{wca}
j^{\mathsf{a}}(z)\,j^{\mathsf{b}}(w)\sim \frac{k\,\delta^{\mathsf{ab}}}{(z-w)^2}+\frac{f^{\mathsf{ab}}_{\:\:\:\:\mathsf{c}}\,j^{\mathsf{c}}}{(z-w)}\,,
\ee
where $k$ is the level of the current algebra and $f^{\mathsf{abc}}$ are the structure constants of $\mathfrak{g}$.\footnote{Note that here the indices $\mathsf{a,b,c}$ run over the dimension of the Lie algebra, and are not to be confused with those of fermionic variables $\eta_{a}$.} We will consider the current algebra to be of level $k=0$ as in~\cite{He:2015zea}, although this assumption can be relaxed for most calculations.

By expanding the function $O(u,\lambda,\tilde{\lambda})$ with respect to the fermionic coordinates, we see that it encodes the radiative degrees of freedom for the gauge field at $\scri_{\C}$. In particular, it follows that
\begin{equation*}
 O(u,\lambda,\tilde{\lambda})=\varphi_{(0,0)}+\cdots+(\eta)^2\cF_{(-2,0)}+\cdots+(\tilde{\eta})^2 \tilde{\cF}_{(0,-2)}+\cdots+(\eta)^2(\tilde{\eta})^2\tilde{\varphi}_{(-2,-2)}\,,
\end{equation*}
with the components being $\mathfrak{g}$-valued functions on bosonic $\scri_{\C}$ whose weight is indicated by the subscripts. The component $\cF_{(-2,0)}$ ($\tilde{\cF}_{(0,-2)}$) is the broadcasting function for a positive (negative) helicity gluon defined by \eqref{broad}, while each of the other components corresponds to the radiative degrees of freedom of the full spectrum of $\cN=4$ super-Yang-Mills theory. For example, the six components coming with an equal number of $\eta$s and $\tilde{\eta}$s represent the scalars.

However, there are additional vertex operators which do not involve the worldsheet current algebra. Roughly, these correspond to deformations of the complex and Hermitian structures of $\scri_{\C}$, and are given by
\begin{equation*}
c\,w\,v\,, \quad c\,\nu^{A}\,v_{A}\,, \quad c\,\tilde{\nu}^{\dot{A}}\,\tilde{v}_{\dot{A}}\,,
\end{equation*}
\be\label{gravo}
c\,\partial u\,g\,, \quad c\,\partial\lambda_{A}\,g^{A}\,, \quad c\,\partial\tilde{\lambda}_{\dot{A}}\,\tilde{g}^{\dot{A}}\,,
\ee
where $v, v_{A}, \tilde{v}_{\dot{A}}, g, g^{A},$ and  $\tilde{g}^{\dot{A}}$ are functions on $\scri_{\C}$ taking values in appropriate powers of $\cL$ and $\tilde{\cL}$ to ensure overall homogeneity. Given their geometric action, it is natural to interpret these vertex operators as \emph{gravitational} perturbations away from the Minkowski vacuum defining $\scri_{\C}$. Indeed, from a holographic perspective we should expect that any model living at null infinity with gauge-theoretic degrees of freedom must be coupled to gravitation in some way. We argue below that \eqref{gravo} actually correspond to a non-unitary theory of gravity. Of course, on the Riemann sphere we may consistently restrict our attention to the gauge-theoretic vertex operators \eqref{gtvo} and to single trace contributions.

\medskip

Given vertex operators, we still require a prescription for computing correlation functions in this CFT. While the $bc$-ghost system provides a measure on the moduli space of $\Sigma$ in the presence of vertex operator insertions, the field $\chi$ from \eqref{S3} has $d+\tilde{d}$ zero modes which must be absorbed. To do this, we note that the composite operator $w\chi$ has conformal weight $(1,0)$ and is un-charged with respect to $\cL,\, \tilde{\cL}$. In the presence of vertex operator insertions at points $\{z_{1},\ldots,z_{n}\}\in\Sigma$, $w\chi$ acquires poles at these insertions. Thus, we define 
\begin{equation*}
 \oint_{z_i} w(z)\,\chi(z)\,,
\end{equation*}
to be the scalar given by the residue of $w\chi$ at $z_i\in\Sigma$.

As an un-charged scalar, this quantity is clearly BRST-closed and has $n-1$ zero modes. Following~\cite{Adamo:2014yya}, we demand that these zero modes also saturate the zero modes of $\chi$ itself, resulting in the constraint
\be\label{zmcounting}
d+\tilde{d}=n-2\,.
\ee
This allows us to eliminate the degree of $\tilde{\cL}$ in terms of $d$ and $n$ in any correlation function, and reproduces the identification between line bundles $\cL\otimes\tilde{\cL}\cong K_{\Sigma}(z_{1}+\cdots+z_{n})$ given by~\cite{Witten:2004cp}.

With this prescription, the correlation function involving $n$ gauge theory vertex operators \eqref{gtvo} is
\begin{multline}\label{corr1}
 \cA_{n,d}=\left\la \prod_{i=1}^{n}c_{i}\, \mathrm{tr}\left(O\; j\right)_{i}\; \prod_{j=1}^{n-3}(b|\mu)_{j}\; \prod_{k=2}^{n}\oint_{\sigma_k} w(\sigma)\,\chi(\sigma)\right\ra \\
 =\left\la c_{1}\, \mathrm{tr}\left(O\; j\right)_{1} c_{2}\, \chi_{2}\,\mathrm{tr}\left(\dot{O}\; j\right)_{2}\; c_{3}\, \chi_{3}\,\mathrm{tr}\left(\dot{O}\; j\right)_3\;\prod_{k=3}^{n}\int_{\Sigma}\chi_{k}\,\mathrm{tr}\left(\dot{O}\;j\right)_k \right\ra\,,
\end{multline}
where a subscript denotes dependence on an insertion point (\textit{e.g.}, $c_1=c(\sigma_1)$) and $\dot{O}=\partial_{u}O$ takes values in $\cO(-1,-1)$ on $\scri_{\C}$.


\section{Kac-Moody Symmetries and Ward Identities}
\label{sec:KMsym}

The correlator \eqref{corr1} involves insertions of the broadcasting function, which encodes the radiative degrees of freedom of (super-)Yang-Mills theory asymptotically. Hence, we expect that any asymptotic symmetries of the gauge field will be realized in the context of our model and have a natural action on correlators.

The most obvious asymptotic symmetry of the gauge theory are global gauge transformations on $\scri_{\C}$. These are implemented in our model by charges
\be\label{globalgt}
\mathcal{J}_{\mathfrak{g}}=\oint \mathrm{tr}\left(\mathsf{T}\,j(\sigma)\right)\,,
\ee
where $\{\mathsf{T}^{\mathsf{a}}\}$ are the generators of the Lie algebra $\mathfrak{g}$ and $j^{\mathsf{a}}(\sigma)$ is the $\mathfrak{g}$-Kac-Moody current. As they are global on $\scri_{\C}$ and holomorphic on the worldsheet, it is obvious that these charges commute with the action and generate exact symmetries of the correlators.

The asymptotic symmetries of Yang-Mills theory are larger than global gauge transformations, though. Strominger showed that the possibility of non-zero color flux through $\scri$ leads to `large' gauge transformations which are constant along the null generators~\cite{Strominger:2013lka}. Such gauge transformations are parametrized by a function on the sphere of null generators which is only locally holomorphic. These are implemented in our model by charges
\be\label{largegt}
\mathcal{J}_{\varepsilon}=\oint \varepsilon(\lambda,\tilde{\lambda})\,\mathrm{tr}\left(\mathsf{T}\,j(\sigma)\right)\,,
\ee
where the function $\varepsilon$ specifies the `large' gauge transformation.\footnote{Of course, this charge actually generates a \emph{complexified} `large' gauge transformation; the real transformation is obtained by restricting to $\scri$.}

Crucially, $\varepsilon$ may have poles, in which case the charge $\mathcal{J}_{\varepsilon}$ is not an exact symmetry due to the residue at those poles. In this case, the contour in \eqref{largegt} is taken around these poles; assuming only simple poles, inserting the charge into the correlator \eqref{corr1} results in a Ward identity:
\begin{multline}\label{WI}
\left< \mathcal{J}_{\varepsilon}\, c_{1}\, \mathrm{tr}\left(O\; j\right)_{1} \cdots \int_{\Sigma}\chi_{n}\,\mathrm{tr}\left(\dot{O}\;j\right)_n \right> \\
= \sum_{w\in\mathcal{S}_n/\mathbb{Z}_n}\sum_{i=1}^{n} \varepsilon(\lambda(\sigma_{w_i}),\tilde{\lambda}(\sigma_{w_i}))\,\mathrm{tr}(\mathsf{T}_{w_1}\cdots[\mathsf{T},\mathsf{T}_{w_i}]\cdots\mathsf{T}_{w_n})\cA_{n,d}\,.
\end{multline}
Here $\mathsf{T}^{\mathsf{a}}_{i}$ is the generator of $\mathfrak{g}$ acting in the representation carried by the vertex operator at insertion $\sigma_{i}\in\Sigma$.  This is closely related to the Ward identity found in~\cite{Strominger:2013lka} for the action of the Kac-Moody current implementing the `large' gauge transformation \eqref{largegt}.

It has been shown that the content of the Ward identity \eqref{WI} should be equivalent to Weinberg's soft gluon theorem~\cite{Strominger:2013lka,He:2015zea}, and this is confirmed in the context of our model with an appropriate choice for the `large' gauge transformation. Indeed, the simplest non-trivial choice for the function $\varepsilon$ is
\be\label{softc}
\varepsilon^{(1)}(\lambda,\tilde{\lambda})=\frac{\la a\,\lambda(\sigma)\ra}{\la a\,s\ra\, \la s\,\lambda(\sigma)\ra}\,,
\ee
where $a_{\alpha}$ is an arbitrary point on $\CP^{1}$, $(\lambda_{s}, \tilde{\lambda}_{s})$ labels the generator of $\scri_{\C}$ associated with the insertion, and we use the standard notation $\la ab\ra=\epsilon_{\alpha\beta} a^{\alpha}b^{\beta}$, $[\tilde{a}\tilde{b}]=\epsilon_{\dot{\alpha}\dot{\beta}}\tilde{a}^{\dot\alpha}\tilde{b}^{\dot\beta}$. Since $\varepsilon^{(1)}$ is homogeneous with respect to $a$ and $\lambda(\sigma)$, and is independent of $\tilde{\lambda}(\sigma)$, the associated charge corresponds to a `holomorphic' Kac-Moody current. Furthermore, $\varepsilon^{(1)}$ has weight $(-2,0)$ with respect to $(\lambda_{s}, \tilde{\lambda}_{s})$, the same as the broadcasting function for a positive helicity gluon.

The single particle states in \eqref{corr1} can be given in the momentum eigenstate representation:
\be\label{momeig}
O^{\mathsf{a}}_{i}=\mathsf{T}^{\mathsf{a}}_{i}\,\int \frac{\d t_{i}\,\d\tilde{t}_{i}}{t_{i} \tilde{t}_{i} \omega_{i}}\,\delta^{2|2}(\lambda_{i}-t_{i}\lambda(\sigma_i))\,\delta^{2|2}(\tilde{\lambda}_{i}-\tilde{t}_{i}\tilde{\lambda}(\sigma_i))\,\e^{t_{i}\tilde{t}_{i}\omega_{i} u(\sigma_i)}\,.
\ee
In a fixed color-ordering, the Ward identity \eqref{WI} then becomes
\begin{multline}\label{soft}
 \frac{a^{\alpha}}{\la a\,s\ra}\left(\left\la \frac{\lambda_{\alpha}(\sigma_1)}{\la s\,\lambda(\sigma_1)\ra} c_{1}\,\mathrm{tr}(O\;j)_{1}\cdots\right\ra -\left\la\frac{\lambda_{\alpha}(\sigma_{n})}{\la s\,\lambda(\sigma_{n})\ra} c_{1}\,\mathrm{tr}(O\;j)_{1}\cdots\right\ra \right) \\
 =\left(\frac{\la a\,1\ra}{\la a\,s\ra\, \la s\,1\ra}-\frac{\la a\,n\ra}{\la a\,s\ra\, \la s\,n\ra}\right)\,\cA_{n,d}=\frac{\la1\,n\ra}{\la s\,1\ra\,\la s\,n\ra}\,\cA_{n,d}\,,
\end{multline}
which is precisely the soft gluon theorem~\cite{Weinberg:1965nx}. So the soft gluon theorem is realized (for a positive helicity soft gluon) by the action of a holomorphic Kac-Moody current generating a `large' gauge transformation at $\scri_{\C}$, which corresponds to the insertion of a soft gluon broadcasting function at $(\lambda_{s},\tilde{\lambda}_{s})\in\CP^{1}\times\CP^{1}$. Analogous to the gravitational model explored in~\cite{Adamo:2014yya}, more general `large' gauge transformations will be related to the insertion of other soft particles in the spectrum of $\cN=4$ super-Yang-Mills or to multiple such soft insertions.

\medskip

Following the discovery of subleading soft factors for gravity~\cite{White:2011yy,Cachazo:2014fwa}, a similar subleading soft factor for gauge theory amplitudes was found~\cite{Casali:2014xpa}. This subleading soft theorem has recently been derived from an asymptotic symmetry perspective for abelian gauge group, where the associated symmetry is not simply a gauge transformation, but also acts as a vector field on the sphere of null generators at $\scri$~\cite{Lysov:2014csa}. It seems natural that the subleading soft theorem is generated by a similar symmetry in the non-abelian setting.  

It is straightforward to write charges generating such rotations on the (complexified) sphere of null generators:
\begin{equation*}
 \mathcal{J}_{V}=\oint V^{\dot{\alpha}}(\lambda,\tilde{\lambda})\tilde{\nu}_{\dot\alpha}(\sigma)\,\mathrm{tr}\left(\mathsf{T}\,j(\sigma)\right)\,,
\end{equation*}
where $V^{\dot\alpha}$ must take values in $\cO(0,-1)$ on $\scri_{\C}$ and have conformal weight $(-1,0)$. There is the possibility for a rotation in the $\lambda$-direction which we have dropped for simplicity. A simple choice for $V^{\dot\alpha}$ is
\be\label{slsoftc}
V^{\dot\alpha}(\lambda,\tilde{\lambda})=\frac{\tilde{\lambda}^{\dot\alpha}_{s}}{w(\sigma)\,\la s\,\lambda(\sigma)\ra}\,,
\ee
and we may consider the Ward identity associated with the action of $\mathcal{J}_{V}$ inside the correlator \eqref{corr1}. Once more, the vector $V^{\dot\alpha}$ has poles which ensure that it is not an exact symmetry of the theory. In order to evaluate the Ward identity when all external states are in the momentum eigenstate representation \eqref{momeig}, we note that the path integral over non-zero modes of $u(\sigma)$ can be performed explicitly, fixing (\textit{c.f.}, \cite{Geyer:2014lca})
\begin{equation*}
 w(\sigma)=\sum_{i=1}^{n}t_{i}\tilde{t}_{i}\omega_{i}\,\frac{\D\sigma_i}{(\sigma\,\sigma_i)}\,\prod_{a=0}^{d+\tilde{d}}\frac{(p_a\,\sigma_i)}{(p_{a}\,\sigma)}\,,
\end{equation*}
where $(\sigma_{i}\,\sigma_{j})$ is the $\SL(2,\C)$-invariant inner product $\epsilon_{\underline{\alpha}\underline{\beta}}\sigma_{i}^{\underline{\alpha}}\sigma_{j}^{\underline{\beta}}$ for homogeneous coordinates on $\Sigma$, $\D\sigma_i=(\sigma_i\,\d\sigma_i)$ is the weight $+2$ holomorphic measure on $\Sigma$, and the $\{p_{a}\}\subset\Sigma$ are an arbitrary collection of $d+\tilde{d}+1$ points. 

Taking the usual contour to pick out the poles, we find (making a choice of color-ordering as before) the Ward identity:
\be\label{WI2}
\left\la \mathcal{J}_{V}\, c_{1}\, \mathrm{tr}(O\; j)_{1} \cdots \int_{\Sigma}\chi_{n}\,\mathrm{tr}(\dot{O}\;j)_n \right\ra = \left(\frac{\tilde{\lambda}^{\dot\alpha}_{s}}{\omega_{1}\,\la s\,1\ra}\frac{\partial}{\partial\tilde{\lambda}^{\dot\alpha}_{1}}-\frac{\tilde{\lambda}^{\dot\alpha}_{s}}{\omega_{n}\,\la s\,n\ra}\frac{\partial}{\partial\tilde{\lambda}^{\dot\alpha}_{n}}\right)\,\cA_{n,d}\,,
\ee
which is the subleading soft factor for a positive helicity gluon inserted between particles $1$ and $n$ in the color ordering~\cite{Casali:2014xpa}. In the abelian case, this is equivalent to the Ward identity used to derive Low's subleading soft theorem~\cite{Lysov:2014csa}. For general gauge group, \eqref{WI2} confirms that the subleading gluon soft factor is related to the action of vector fields on the conformal two-sphere explicitly.


\section{Scattering Amplitudes}
\label{sec:amps}

At this point, we have only considered correlators involving gauge-theoretic vertex operators \eqref{gtvo} encoding the broadcasting data of the gauge field. On the Riemann sphere this restriction is consistent and corresponds to isolating single trace contributions in the color structure. The Ward identity \eqref{WI} establishes that charges implementing the action of asymptotic `large' gauge transformations act on these correlators in a natural way, implying the soft gluon theorem. This extends to other charges, acting as rotations on the space of generators of $\scri_{\C}$, which give the subleading soft theorem. We now turn to the actual evaluation of \eqref{corr1}, as well as a discussion of the other states in the theory given by vertex operators \eqref{gravo}.

To evaluate \eqref{corr1}, we consider all vertex operator insertions to be given by the momentum eigenstates \eqref{momeig}. The only non-trivial Wick contractions between the vertex operators are in the worldsheet current algebra; using \eqref{wca} and choosing a specific color-ordering this leads to the usual Parke-Taylor factor:
\begin{equation*}
 \left\la \prod_{i=1}^{n} j^{\mathsf{a}_i}(\sigma_i)\right\ra=\mathrm{tr}\left(\mathsf{T}^{\mathsf{a}_1}\cdots\mathsf{T}^{\mathsf{a}_n}\right)\,\prod_{i=1}^{n} \frac{\D\sigma_i}{(\sigma_{i}\,\sigma_{i+1})}\,.
\end{equation*}
The remainder of the correlator is given by the zero mode integrals for the various worldsheet fields, with the result:
\begin{multline}\label{amp1}
\cA_{n,d}=\mathrm{tr}\left(\mathsf{T}^{\mathsf{a}_1}\cdots\mathsf{T}^{\mathsf{a}_n}\right) \int \frac{\prod_{r=0}^{d}\d^{2|2}\lambda^{0}_{r}\,\prod_{s=0}^{\tilde{d}}\d^{2|2}\tilde{\lambda}^{0}_{s}}{\mathrm{vol}(\C^{*}\times\C^{*})}\frac{|\sigma_{1}\sigma_{2}\sigma_{3}|}{\D\sigma_1 \D\sigma_2 \D\sigma_3}\frac{|\sigma_{2}\cdots\sigma_{n}|}{t_{1}\tilde{t}_{1}\omega_{1}} \\
\prod_{a=0}^{d+\tilde{d}}\delta\left(\sum_{i=1}^{n} t_{i}\tilde{t}_{i}\omega_{i}\,\mathfrak{s}_{a}(\sigma_i)\right)\,\prod_{i=1}^{n}\frac{\D\sigma_{i}}{(\sigma_{i}\,\sigma_{i+1})}\d t_{i}\,\d\tilde{t}_{i}\,\delta^{2|2}(\lambda_{i}-t_{i}\lambda(\sigma_i))\,\delta^{2|2}(\tilde{\lambda}_{i}-\tilde{t}_{i}\tilde{\lambda}(\sigma_i)) 
\end{multline}
In this expression, the measure in the first line is over the zero modes of the maps $\lambda_{A}(\sigma),\,\tilde{\lambda}_{\dot{A}}(\sigma):\Sigma\rightarrow\CP^{1|2}$, the two $\C^{*}$-freedoms are associated with the scalings of $\cL$ and $\tilde{\cL}$, and the Vandermonde determinants
\begin{equation*}
 \frac{|\sigma_{1}\sigma_{2}\sigma_{3}|}{\D\sigma_1\, \D\sigma_2\, \D\sigma_3}=\frac{(\sigma_1\,\sigma_2)(\sigma_2\,\sigma_3)(\sigma_3\,\sigma_1)}{\D\sigma_1 \D\sigma_2 \D\sigma_3}\,, \qquad |\sigma_{2}\cdots\sigma_{n}|=\prod_{2\leq i<j\leq n} (\sigma_{i}\,\sigma_{j})\,,
\end{equation*}
are given by the $3$ zero modes of $c$ and $n-1$ zero modes of $\chi$, respectively. The delta functions in the second line are from momentum eigenstate insertions and the integral over $u(\sigma)$ zero modes:
\begin{equation*}
 \int \prod_{a=0}^{d+\tilde{d}}\d u^{0}_{a}\,\exp\left[\sum_{i=1}^{n}t_{i}\tilde{t}_{i}\omega_{i}\,u(\sigma_{i})\right] = \prod_{a=0}^{d+\tilde{d}}\delta\left(\sum_{i=1}^{n}t_{i}\tilde{t}_{i}\omega_{i}\,\mathfrak{s}_{a}(\sigma_i)\right)\,,
\end{equation*}
where $\{\mathfrak{s}_{a}\}$ form a basis of $H^{0}(\Sigma, \cL\otimes\tilde{\cL})$. 

This expression can be manipulated into a more recognizable form by using the identity (\textit{c.f.}, \cite{Roiban:2004yf,Witten:2004cp,Cachazo:2013zc})
\begin{equation*}
 \prod_{a=0}^{d+\tilde{d}}\delta\left(\sum_{i=1}^{n}t_{i}\tilde{t}_{i}\omega_{i}\,\mathfrak{s}_{a}(\sigma_i)\right)=\frac{1}{|\sigma_1 \cdots\sigma_{n}|}\int \d r \prod_{i=1}^{n}\delta\left(t_{i}\tilde{t}_{i}\omega_{i}-\frac{r}{\prod_{j\neq i}(\sigma_i\,\sigma_{j})}\right)\,.
\end{equation*}
The result,
\begin{multline}\label{amp2}
 \cA_{n,d}=\mathrm{tr}\left(\mathsf{T}^{\mathsf{a}_1}\cdots\mathsf{T}^{\mathsf{a}_n}\right) \int \frac{\prod_{r=0}^{d}\d^{2|2}\lambda^{0}_{r}\,\prod_{s=0}^{\tilde{d}}\d^{2|2}\tilde{\lambda}^{0}_{s}}{\mathrm{vol}(\SL(2,\C)\times\C^{*}\times\C^{*})}\, \frac{\d r}{r}\prod_{i=1}^{n}\delta\left(t_{i}\tilde{t}_{i}\omega_{i}-\frac{r}{\prod_{j\neq i}(\sigma_i\,\sigma_{j})}\right) \\
 \times \frac{\D\sigma_{i}}{(\sigma_{i}\,\sigma_{i+1})}\,\d t_{i}\,\d\tilde{t}_{i}\,\delta^{2|2}(\lambda_{i}-t_{i}\lambda(\sigma_i))\,\delta^{2|2}(\tilde{\lambda}_{i}-\tilde{t}_{i}\tilde{\lambda}(\sigma_i))\,
\end{multline}
is equal to the parity-invariant form of the Roiban-Spradlin-Volovich~\cite{Roiban:2004yf} expression for the classical S-matrix of $\cN=4$ super-Yang-Mills given by Witten~\cite{Witten:2004cp}.  Thus, the correlators \eqref{corr1} of our model reproduce all tree-level amplitudes of gauge theory, confirming the interpretation of the Ward identities in the previous section.

\medskip

Now let us consider the gravitational degrees of freedom in the model, corresponding to the vertex operators \eqref{gravo}. While consistently omitted from single trace gauge theory interactions at tree-level, these states can mediate multi-trace tree-level amplitudes in the gauge theory and would also run in any amplitudes computed to higher-order in perturbation theory (though we do not address such loop-level questions here). We claim that these degrees of freedom correspond to a non-unitary theory of gravity with fourth-order equations of motion.

The non-unitary nature of the gravitational degrees of freedom is manifest by considering a multi-trace correlator where all external states are given by gauge-theoretic vertex operators \eqref{gtvo}. Since the only non-trivial Wick contractions between these operator insertions are in the worldsheet current algebra, the arguments of~\cite{Adamo:2013tca} ensure that this double trace is mediated by gravitational degrees of freedom with fourth-order equations of motion. As an explicit example, consider a double trace correlator of $n$ external states, with $n_1$ in one trace and $n_2$ in the other ($n_1+n_2=n$). It is straightforward to show that such a correlator can be written (for momentum eigenstates) as:
\begin{multline*}
 \mathrm{tr}\left(\mathsf{T}^{\mathsf{a}_1}\cdots\mathsf{T}^{\mathsf{a}_{n_{1}}}\right)\,\mathrm{tr}\left(\mathsf{T}^{\mathsf{b}_1}\cdots\mathsf{T}^{\mathsf{b}_{n_{2}}}\right)\int \frac{\prod_{r=0}^{d}\d^{2|2}\lambda^{0}_{r}\,\prod_{s=0}^{\tilde{d}}\d^{2|2}\tilde{\lambda}^{0}_{s}}{\mathrm{vol}(\SL(2,\C)\times\C^{*}\times\C^{*})}\, \frac{\d r}{r} \\
 \times \prod_{j=1}^{n_{1}}\frac{\D\sigma_{j}}{(\sigma_{j}\,\sigma_{j+1})}\,\d t_{j}\,\d\tilde{t}_{j}\,\delta\left(t_{j}\tilde{t}_{j}\omega_{j}-\frac{r}{\prod_{l\neq j}(\sigma_j\,\sigma_{l})}\right)\,\delta^{2|2}(\lambda_{j}-t_{j}\lambda(\sigma_j))\,\delta^{2|2}(\tilde{\lambda}_{j}-\tilde{t}_{j}\tilde{\lambda}(\sigma_j)) \\
 \times \prod_{k=1}^{n_{2}}\frac{\D\sigma_{k}}{(\sigma_{k}\,\sigma_{k+1})}\,\d t_{k}\,\d\tilde{t}_{k}\,\delta\left(t_{k}\tilde{t}_{k}\omega_{k}-\frac{r}{\prod_{m\neq k}(\sigma_k\,\sigma_{m})}\right)\,\delta^{2|2}(\lambda_{k}-t_{k}\lambda(\sigma_k))\,\delta^{2|2}(\tilde{\lambda}_{k}-\tilde{t}_{k}\tilde{\lambda}(\sigma_k))\,.
\end{multline*}

Consider the limit where this correlator factorizes without splitting the color traces: the worldsheet $\Sigma$ degenerates into two Riemann spheres $\Sigma_1$ and $\Sigma_2$ attached at a node, with vertex operators $1,\ldots,n_1$ on $\Sigma_1$ and $1,\ldots, n_2$ on $\Sigma_2$. In this limit the only states flowing through the factorization channel are gravitational. The worldsheet can be modeled on a conic in $\CP^{2}$ with a (complex) parameter $q$ controlling the degeneration; in the $q\rightarrow 0$ limit, $\Sigma$ factorizes into $\Sigma_1\cup\Sigma_2$, with the marked points $\sigma_{x}\in\Sigma_1$ and $\sigma_{y}\in\Sigma_{2}$ identified at the node.

Standard arguments (\textit{c.f.}, Appendix C of~\cite{Adamo:2013tca}) show that in the $q\rightarrow0$ limit, the portion of the correlator encoding the trace structure factorizes as
\begin{multline*}
 \frac{1}{\mathrm{vol}\;\SL(2,\C)}\prod_{j=1}^{n_{1}}\frac{\D\sigma_{j}}{(\sigma_{j}\,\sigma_{j+1})}\prod_{k=1}^{n_{2}}\frac{\D\sigma_{k}}{(\sigma_{k}\,\sigma_{k+1})} \\ \rightarrow \frac{\d q}{q^2}\,\left(\frac{\D\sigma_{x}}{\mathrm{vol}\;\SL(2,\C)}\prod_{j=1}^{n_{1}}\frac{\D\sigma_{j}}{(\sigma_{j}\,\sigma_{j+1})}\right)\,\left(\frac{\D\sigma_{y}}{\mathrm{vol}\;\SL(2,\C)}\prod_{j=1}^{n_{2}}\frac{\D\sigma_{k}}{(\sigma_{k}\,\sigma_{k+1})}\right)\,+O(q^{-1})\,.
\end{multline*}
Due to the $\cN=4$ supersymmetry in play, all other parts of the correlator are homogeneous, introducing no additional powers of $q$. The various delta functions can be used to show that in the factorization limit, $q$ scales as the square of the total momentum flowing through the channel. Thus, the presence of a double pole in $q$ indicates a momentum space propagator of the form $p^{-4}$, as expected for a theory with fourth-order equations of motion.

\medskip

While this factorization argument confirms that the gravitational vertex operators of our model do not correspond to Einstein gravity, we can be more precise by considering correlators of the operators \eqref{gravo} themselves. We argue that the gravitational interactions are consistent with a particular \emph{non-minimal} $\cN=4$ conformal supergravity (\textit{c.f.}, \cite{Fradkin:1985am}) arising in the twistor-strings of Witten and Berkovits~\cite{Witten:2003nn,Berkovits:2004hg,Berkovits:2004jj}.\footnote{`Non-minimal' refers to the presence of interaction terms between scalars and two Weyl tensors in the space-time action~\cite{Fradkin:1982xc}. The existence of this non-minimal theory at the quantum level is questionable due to $\SU(4)$ axial anomaly calculations~\cite{Romer:1985yg}, but we confine our attention to the semi-classical observables.} Of course, conformal invariance is explicitly broken in our model at the level of the target space: $\scri_{\C}$ is topologically distinct from the conformal boundary of (anti-)de Sitter space, which is a conformally equivalent bulk space-time. Minkowski space \emph{is} a vacuum solution to the conformal gravity equations of motion, though, so the vertex operators \eqref{gravo} can be thought of as linearized perturbations around this fixed background conformal structure.

Low-point amplitudes of non-minimal conformal supergravity have been calculated in the context of twistor-string theory~\cite{Berkovits:2004jj,Ahn:2005es,Dolan:2008gc,Adamo:2012nn}, and the structure of the vertex operators makes it clear that our model will reproduce those amplitudes in a parity-symmetric form, analogous to the gauge theory calculation above. For instance, the $n=3$, $d=0$ correlator
\begin{multline*}
\left\la\prod_{i=1}^{3}c_{i}\,\nu_{i}^{A}\,v_{i\,A}\,\prod_{j=2,3}\oint_{\sigma_j} w(\sigma)\,\chi(\sigma)\right\ra \\
=\int\frac{\d^{2|2}\lambda^{0}\,\prod_{s=0,1}\d^{2|2}\tilde{\lambda}^{0}_{s}\prod_{t=0}^{2}\d u^{0}_{t}}{\mathrm{vol}\left(\C^{*}\times\C^{*}\right)} \left(\frac{\partial v_{1\,B}}{\partial\lambda_{A}} \frac{\partial \dot{v}_{2\,C}}{\partial\lambda_{B}} \frac{\partial \dot{v}_{3\,A}}{\partial\lambda_{C}}-\frac{\partial v_{1\,B}}{\partial\lambda_{C}} \frac{\partial \dot{v}_{2\,C}}{\partial\lambda_{A}} \frac{\partial \dot{v}_{3\,A}}{\partial\lambda_{B}}\right)\,,
\end{multline*}
is non-vanishing, and corresponds to a cubic interaction between two conformal gravitons and a conformal scalar in Minkowski space. This interaction is forbidden in minimal $\cN=4$ conformal supergravity by a global $\SU(1,1)$ symmetry acting on the conformal scalar~\cite{Bergshoeff:1980is,Fradkin:1985am}.

Another test of non-minimality is given by embedding Einstein degrees of freedom inside the gravitational vertex operators. Fixing a conformal structure to perform this embedding corresponds to picking a cosmological constant, $\Lambda$; in minimal conformal supergravity, correlators of the embedded Einstein operators will be $O(\Lambda)$ polynomials~\cite{Maldacena:2011mk,Adamo:2013tja}. In our model, such an embedding can be given by taking linear combinations of \eqref{gravo}:
\begin{equation*}
\nu^{A}\,v_{A}+\tilde{\nu}^{\dot{A}}\,\tilde{v}_{\dot{A}}\rightarrow \Lambda \nu^{A}\frac{\partial h}{\partial\lambda_{A}} + \tilde{\nu}^{\dot{A}}\frac{\partial\tilde{h}}{\partial\tilde{\lambda}_{\dot{A}}}\,, \qquad \partial\lambda_{A}\,g^{A}+\partial\tilde{\lambda}_{\dot{A}}\,\tilde{g}^{\dot{A}}\rightarrow \la\partial\lambda\,\lambda\ra\,G+\Lambda\,[\partial\tilde{\lambda}\,\tilde{\lambda}]\,\tilde{G}\,,
\end{equation*}
where $h$, $\tilde{h}$, $G$, and $\tilde{G}$ encode Einstein degrees of freedom. For example, in the expansion
\begin{equation*}
 h=f_{(-2,0)}+\cdots+(\eta)^{2} N_{(-4,0)}+\cdots+(\tilde{\eta})^{2} \phi_{(-2,-2)}+\cdots +(\eta)^{2}(\tilde{\eta})^{2} \tilde{f}_{(-4,-2)}\, 
\end{equation*}
the component $N_{(-4,0)}$ is the news function of a positive helicity Einstein graviton, while $f_{(-2,0)}$, $\tilde{f}_{(-4,-2)}$ and $\phi_{(-2,-2)}$ encode the radiative degrees of freedom for photons and scalars in $\cN=4$ supergravity. It is easy to see that correlators of these operators will have an $O(\Lambda^0)$ piece, indicating non-minimal structure.\footnote{The $\Lambda\rightarrow0$ limit of certain correlators also matches those produced by the non-minimal Berkovits-Witten model at arbitrary $n$ and $d$~\cite{Adamo:2012xe}, though there are other correlators whose interpretation is less clear.}

While this hardly suffices to establish that the gravitational interactions of our model are \emph{equivalent} to non-minimal conformal supergravity, it does seem to indicate that this theory is at least a subsector of our model (presented in a conformally broken target space framework). Combined with other obvious similarities to the Berkovits-Witten twistor-string, this suggests that there could be a transcription (in some sense) of our model from $\scri_{\C}$ to twistor space.    



\section{Discussion}
\label{sec:con}

In this work we have constructed a two-dimensional model with target space the complexified null boundary of Minkowski space, which contains in its spectrum the scattering states of $\mathcal{N}=4$ super-Yang-Mills and is naturally acted upon by the asymptotic gauge symmetries. Specific correlators of this CFT compute the tree-level S-matrix of gauge theory when restricted to single trace. We identified the charges that generate `large' gauge transformations at $\scri$ and showed that their insertion in a correlator with a specific parameter reproduces the soft factor for a gluon of vanishing energy. We also identified the charge that generates the subleading soft factor, which is a rotation on the space of null generators accompanied by a gauge transformation. We saw that, as expected from holography, the model contains states corresponding to gravitational degrees of freedom; we argued that these are non-unitary in nature.

It is clear that this model shares many features of the Berkovits-Witten twistor-string, just as the gravitational model of~\cite{Adamo:2014yya} shares many features of Skinner's twistor-string for $\cN=8$ supergravity~\cite{Skinner:2013xp}. It would be interesting to explicitly find the relation between the models on $\scri_{\C}$ and their twistorial analogues, as well as their relation to the four-dimensional ambitwistor string~\cite{Geyer:2014fka}.

While the models considered in this paper and in~\cite{Adamo:2014yya} are intrinsically four-dimensional, the soft gluon (and graviton) theorems hold for semi-classical scattering in \emph{any} dimension, and the same is true for their subleading corrections~\cite{Schwab:2014xua,Afkhami-Jeddi:2014fia}. It is also clear that the connection between these soft theorems and asymptotic symmetries persists in dimensions beyond four~\cite{Kapec:2014zla,Kapec:2015vwa}. The ambitwistor formalism~\cite{Mason:2013sva} extends to arbitrary dimension, and can be used to derive leading and subleading soft theorems (as well as their quantum corrections)~\cite{Lipstein:2015rxa,Geyer:2014lca}. However, the presentation of the asymptotic data in the ambitwistor models is not in terms of broadcasting (or news) functions. It would be interesting to know if higher-dimensional versions of our model (where asymptotic data is manifestly given in terms of radiative modes) exist.

A final outstanding issue is to determine the complete boundary theory for which our model is a perturbative worldsheet description. As in~\cite{Adamo:2014yya}, our model only describes the classical perturbative regime of such a theory. While the field theory (possibly non-local) on $\scri_{\C}$ corresponding to our model would still be classical in nature, it may nevertheless contain important clues about flat space holography. We hope to address this in future works.   

\acknowledgments

We thank Arthur Lipstein, Lionel Mason, and David Skinner for useful conversations and comments. The work of TA is supported by a Title A Research Fellowship at St. John's College, Cambridge. The work of EC is supported in part by the Cambridge Commonwealth, European and International Trust.

\bibliography{scri_2}
\bibliographystyle{JHEP}

\end{document}